\documentclass[conference]{IEEEtran}
\IEEEoverridecommandlockouts
\usepackage{cite}
\usepackage{amsmath, amsfonts, amssymb, amsthm}
\usepackage{algorithm}
\usepackage{algorithmic}
\usepackage{graphicx}
\usepackage{textcomp}
\usepackage{xcolor}
\usepackage{subfigure}

\def\BibTeX{{\rm B\kern-.05em{\sc i\kern-.025em b}\kern-.08em
    T\kern-.1667em\lower.7ex\hbox{E}\kern-.125emX}}
\begin{document}

\title{Nash Soft Actor-Critic LEO Satellite Handover Management Algorithm for Flying Vehicles}

	\author{Jinxuan Chen\textsuperscript{*}, Mustafa Ozger\textsuperscript{*}, Cicek Cavdar\textsuperscript{*}
		\\
		\textsuperscript{*}School of Electrical Engineering and Computer Science, KTH Royal Institute of Technology, Sweden
		\\ 
		Email: \textsuperscript{*}\{jinxuanc, ozger, cavdar\}@kth.se
	}	
	\maketitle

\begin{abstract}
Compared with the terrestrial networks (TN), which can only support limited coverage areas, low-earth orbit (LEO) satellites can provide seamless global coverage and high survivability in case of emergencies. Nevertheless, the swift movement of the LEO satellites poses a challenge: frequent handovers are inevitable, compromising the quality of service (QoS) of users and leading to discontinuous connectivity. Moreover, considering LEO satellite connectivity for different flying vehicles (FVs) when coexisting with ground terminals, an efficient satellite handover decision control and mobility management strategy is required to reduce the number of handovers and allocate resources that align with different users' requirements. In this paper, a novel distributed satellite handover strategy based on Multi-Agent Reinforcement Learning (MARL) and game theory named Nash-SAC has been proposed to solve these problems. From the simulation results, the Nash-SAC-based handover strategy can effectively reduce the handovers by over 16\% and the blocking rate by over 18\%, outperforming local benchmarks such as traditional Q-learning. It also greatly improves the network utility used to quantify the performance of the whole system by up to 48\% and caters to different users’ requirements, providing reliable and robust connectivity for both FVs and ground terminals.
\end{abstract}


\section{Introduction}
With the increasing demand for high-quality services from the aviation industry, the concept of Future Aerial Communications (FACOM) was introduced as the connectivity ecosystem that incorporates all these looming aerial connectivity use cases and their potential connectivity solutions \cite{b1}. In aerial use cases, flying vehicles (FVs) such as electrical vertical take-off and landing (eVTOL) vehicles and unmanned aerial vehicles (UAVs) require continuous connectivity even when they fly over remote regions that are not covered with the ground communication infrastructure, i.e., lakes, oceans, and mountainous areas. Satellite communication (SC), as one of the most promising technologies, can provide global communication connectivity to FVs. Low-earth orbit (LEO) satellites located at an altitude between 100-2000 km offer low latency and available high bandwidth services with ubiquitous connectivity and their energy consumption is also low compared with geosynchronous-earth orbit (GEO) and medium-earth orbit (MEO) satellites \cite{b2}.

However, the rapid handover rate of the LEO satellite network due to the high speed of the LEO satellites leads to discontinuous connectivity and imperfect overall network availability \cite{b3}. The satellite handover is the process by which a satellite communication link is transferred from one satellite to another as the user moves across the coverage area of the satellites \cite{b3}. During the handover process, the satellite needs to use extra power to establish a connection with the new satellite, especially for the transmission of measurement reports by users \cite{b4}, which can result in increased power consumption and discontinuous connectivity. As the communication link is handed over from one satellite to another, there can be a brief disruption in the signal which may cause dropped calls, interrupted data transmissions, or degraded signal quality, leading to a poor user experience \cite{b5}. 

To address the potential issues caused by satellite handovers, many handover strategies have been proposed. In \cite{b6}, the Dynamic Doppler-Based Handover Prioritization (DDBHP) scheme is introduced to forecast handover load, thereby improving the management of handover requests, but with only a focus on optimizing channel utilization and bandwidth efficiency. In \cite{b7}, two distinct algorithms are proposed to cater to varying circumstances to enhance call quality, mitigate network congestion, and minimize the number of satellite handovers. However, this paper only considers the ground terminals, and the results use the SNR indicator neglecting the influence of the interference signals which is imperfect. In \cite{b8}, an entropy-based multi-objective optimization handover strategy is proposed when considering four key handover elements including interference signals, to achieve better performance, but it was the local optimization solutions with the instantaneous information.  In \cite{b9}, the satellite handover problem is transformed in the LEO satellite network into the multi-agent reinforcement learning (MARL) problem, which is solved by a distributed Q-learning algorithm including Boltzmann exploration and $\epsilon$-greedy policy. However, the method in this paper sacrifices the blocking rate for the reduction of handovers which leads to network congestion. In \cite{b10}, deep reinforcement learning is leveraged to facilitate handover decisions, taking into account several factors, including link quality and network congestion. Although the proposed scheme decreases the number of handovers, the average carrier-to-noise ratio (CNR) and interference-to-noise ratio (INR) are not optimized. 

Most of the existing studies only consider one specific handover criterion such as maximum instantaneous signal strength \cite{b11}, number of available channels \cite{b12}, and elevation angle \cite{b13}, which is not comprehensive for the handover scenario. Moreover, they are not sufficient given the high dimensionality and flexibility of the systems which only consider satellites and UE with small quantities, and they are only focused on the connectivity between the satellites and ground user with inadequate consideration \cite{b6} - \cite{b13}. Last but not least, the current literature has incomplete performance in terms of the number of handovers where there is still significant room for improvement. In this paper, a novel satellite handover strategy based on MARL and game theory, named Nash-SAC, has been proposed to solve these problems. The main contributions are as follows: 
\begin{itemize}
\item The high-dimensionality user-satellite network is constructed including the LEO constellation from ephemeris data and different types of users, including FVs such as aircraft, UAVs, eVTOLs, and ground terminals.
\item Comprehensive criteria including the remaining visible time, the signal quality CNR, the channel interference INR, and the available idle channel are considered to form the two mathematical optimization problems, the traditional low handover model to minimize the number of handovers and the network utilization model to maximize the network utility.
\item The performance of the proposed satellite handover strategy based on Nash-SAC achieves the best performance in terms of satellite handovers, the blocking rate, and the network utility of the whole system compared with different benchmarks, and caters to different users' requirements, providing reliable and robust connectivity for both FVs and ground terminals.
\end{itemize}
The remainder of this paper is organized as follows. Section~II introduces the design of the handover management strategy including the system model, the mathematical optimization problem, and the Nash-SAC algorithm. In Section~III, parameters are set and the performance of the proposed scheme is evaluated with different benchmarks. Finally, conclusions are drawn in Section IV.
\section{Design of Handover Management Strategy}

\subsection{System model}
Considering the satellite handover scenario with Earth-moving cells (EMC) in the time period $T$, the architecture of the LEO satellite network is shown in Fig. 1, including the LEO satellites in continuous motion (Walker Star constellation) and different types of FVs and ground terminals. The typical LEO satellite network has $N$ LEO satellites, which is denoted by $ \mathcal{N} = \{ 1,2,...,N\} $, and provides seamless service for $K$ users including ${K_1}$ ground terminals and ${K_2}$ FVs which can be denoted by $\mathcal{K} = \{ 1,2,...,{K_1},{K_1} + 1,...,{K_1} + {K_2}\} $. Then, we divide the time period $T$ into $U$ sections, which can be denoted by $\mathcal{U}  = \{ {t_1},{t_2},...,{t_U}\}$, in which the satellite coverage graph is considered to remain unchanged in each section \cite{b9}, \cite{b10}. The user routinely gathers information about the network, specifically at the beginning of each interval, and then decides if a handover is necessary based on the strategy.
\begin{figure}[htbp]
\centerline{\includegraphics[width=3in]{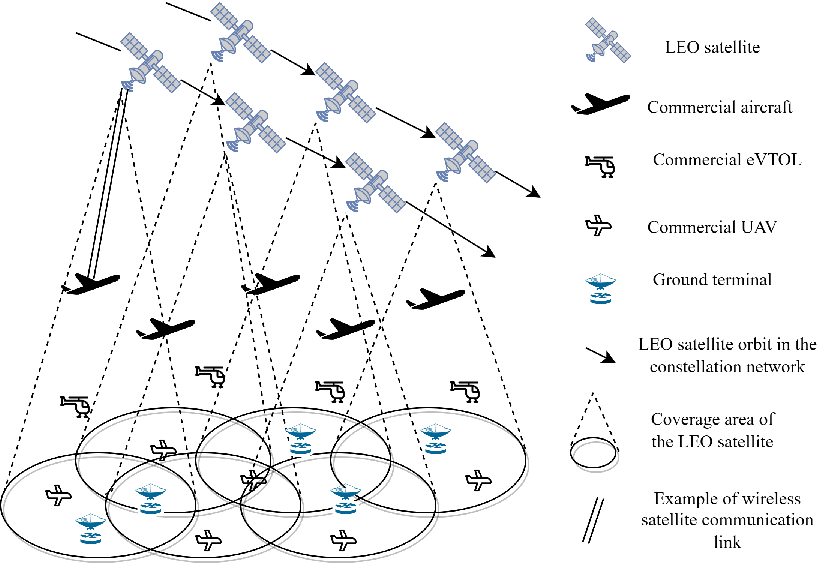}}
\caption{The architecture of the satellite network and the users, including the FVs and the ground terminals.}
\label{fig}
\end{figure}
During each section, it is assumed that the user is cognizant of its precise location through the use of the Global Positioning System (GPS). Furthermore, due to the predictable trajectory of LEO satellites, it can identify all the satellites covering its location \cite{b14}. However, it can only be served by one satellite it chooses. Therefore, two indicators for coverage and association, respectively, are introduced as
\begin{equation}
c_{k,n}^t = \left\{ {\begin{array}{*{20}{c}}
{{\rm{ }}1,\theta _{k,n}^t \ge {\theta _{threshold}},\forall n \in \mathcal{N},\forall k \in \mathcal{K},\forall t \in \mathcal{U}}\\
{0,{\rm{ else}}{\rm{.}}}
\end{array}} \right.
\end{equation}
\begin{equation}
x_{k,n}^t = \left\{ {\begin{array}{*{20}{c}}
{{\rm{ }}1,{\rm{ if ~user }}~k{\rm{ ~is ~served ~by ~satellite }}~n{\rm{ ~in ~section }}~t}\\
{0,{\rm{ else}}{\rm{.}}}
\end{array}} \right.
\end{equation}
where $\theta _{k,n}^t$ represents the elevation angle.

\subsection{Problem formulation} 
Considering the downlink (DL) communication, CNR is used in wireless networking to evaluate the quality of a received signal, which can be expressed as
\begin{equation}
\resizebox{1\hsize}{!}{$CN{R_{k,n}}= \left\{ \begin{array}{l}
10{\log _{10}}(\frac{{{p_t}{g_{i,1}}{g_{i,2}}}}{{{k_z}{T_{e,i}{B}}{\zeta_{k,n}}}}),i \in \{1,2,3,4\} {\rm{ , if }}~c_{k,n}^t = 1\\
0,{\rm{ else}}
\end{array} \right.$}
\end{equation}
where ${{p_t}}$ is the transmit power,  ${{k_z}}$ is the Boltzmann’s constant, ${{T_{e,i}}}$ is the noise temperature of the receiving system, $B$ is the transmit bandwidth, $\zeta_{k,n}$ is the free space path loss of the DL, $i \in \{ 1,2,3,4\}$ represents types of users, i.e., airplanes, eVTOLs, UAVs, and ground terminals, respectively, and ${{g_{i,1}}}$ and ${{g_{i,2}}}$ are the transmitting and receiving parabolic antenna gain for user type of $i$.

In SC, there may be interfering signals caused by the satellites which cover the users but do not provide the service for them. The high INR indicates that interfering signals are significantly contributing to the overall noise in the system, which can degrade the performance and reliability of the communication link. It can be expressed as
\begin{equation}
IN{R_{k,n}} = \left\{ {\begin{array}{*{20}{l}}
{10{{\log }_{10}}\sum\limits_{n = 1}^N {\frac{{{p_t}{g_{i,1}}{g_{i,2}}}}{{{k_z}{T_{e,i}}{B}{\zeta_{k,n}}}}} , {\rm{if}}\;c_{k,n}^t = 1,x_{k,n}^t \ne 1}\\
{0,{\rm{else}}}.
\end{array}} \right.
\end{equation}

Subsequently, we can obtain the Carrier-to-Interference-plus-Noise Ratio (CINR), which can be expressed as
\begin{equation}
\vartheta  = \frac{S}{{{I_N} + {I_F}}} = \frac{{\frac{S}{{{I_F}}}}}{{\frac{{{I_N}}}{{{I_F}}} + 1}} = \frac{{CNR}}{{INR + 1}},
\end{equation}
where $S$ is the strength of the received useful signal, ${I_N}$ and ${I_F}$ represent the strength of received interference and noise signals, respectively. 
The satellite handover management problem can be formulated as a mathematical optimization task. In this context, the primary aim of the conventional low handover model is to minimize the number of handovers while simultaneously alleviating network congestion, which can be expressed as follows:
\begin{align} \label{min1}
\min_{x_{k,n}^t} \quad & \sum\limits_{k = 0}^K {{HO_k}} \\
\textrm{s.t.} \quad & x_{k,n}^t \in \{ 0,1\} ,\forall n \in \mathcal{N},\forall k \in \mathcal{K},\forall t \in \mathcal{U}, \tag{6a}\\
     & \sum\limits_{k = 0}^K {x_{k,n}^t \le L,} \forall n \in \mathcal{N},\forall t \in \mathcal{U}, \tag{6b}\\
     & \sum\limits_{n = 0}^N {x_{k,n}^t = 1,} \forall k \in \mathcal{K},\forall t \in \mathcal{U},\tag{6c}\\
     & \vartheta _{k,n}^t > {\vartheta _{threshold}},\forall n \in \mathcal{N},\forall k \in \mathcal{K},\forall t \in \mathcal{U}, \tag{6d}
\end{align}
where ${HO}_k$ is the total number of handovers for the user $k$ during the time period $T$. The first constraint in (6a) indicates if the user $k$ is served by satellite $n$ at time $t$. The second constraint in (6b) indicates that each satellite is allocated certain $L$ channels, thereby reasonable management strategy can improve channel utilization efficiency and avoid network congestion. The two constraints in (6c) and (6d) indicate that each user will only be served by one satellite and the minimum CINR threshold needs to be met for reliable communication, respectively.

In some cases, both a low number of handovers and high QoS are significant in providing users with an ultra-reliable and high-quality communication experience. The network utility \cite{b15} is introduced to quantify the overall performance and efficiency delivered by the satellite network to its users. Hence, the second mathematical problem is the multi-objective optimization problem, which is aimed at maximizing the network utility, capturing handovers, CINR, and blocking rate according to different QoS requirements. This problem is formulated as follows:
\begin{align}  \label{min2}
\max_{x_{k,n}^t} \quad & \psi  = {\sum\limits_{k = 1}^K {({-w_1} \cdot H{O_k} + {w_2} \cdot \vartheta  - {w_3} \cdot \varsigma } )}\\
\textrm{s.t.} \quad & x_{k,n}^t \in \{ 0,1\} ,\forall n \in \mathcal{N},\forall k \in \mathcal{K},\forall t \in \mathcal{U}, \tag{7a}\\
& \sum\limits_{n = 0}^N {x_{k,n}^t = 1,} \forall k \in \mathcal{K},\forall t \in \mathcal{U}, \tag{7b}\\
& \vartheta _{k,n}^t > {\vartheta _{threshold}},\forall n \in \mathcal{N},\forall k \in \mathcal{K},\forall t \in \mathcal{U}, \tag{7c}\\
& \varsigma  ={\sum\limits_{t = 1}^U {\sum\limits_{n = 1}^N {\max (0,(\sum\limits_{k = 0}^K {x_{k,n}^t - L))} } } }, \tag{7d}
\end{align}
where $w_1,w_2,w_3$ are the weights and $\varsigma$ is the blocking rate, which represents the occurrence of satellite overload.

\subsection{Nash-SAC}
The two mathematical problems formulated above are the combinatorial integer optimization problems, which both are NP-hard in general. To use RL algorithms, we need to define the agent, state, action, and reward function, which are indicated as follows.
\begin{itemize}
\item[$\bullet$]Agent: User $ k\in \mathcal{K}$ including ground terminals and FVs. They take action at the beginning of the intervals, which leads to the state transition.
\item[$\bullet$]State: $s_k^t$ represents the state of $k-th$ agent at section $t$, which is expressed as:
\begin{equation}
s_k^t = \left[ {\begin{array}{*{20}{c}}
{c_{k,1}^t}&{c_{k,2}^t}&{...}&{c_{k,n}^t}\\
{z_{k,1}^t}&{z_{k,2}^t}&{...}&{z_{k,n}^t}\\
{\vartheta _{k,1}^t}&{\vartheta _{k,2}^t}&{...}&{\vartheta _{k,n}^t}\\
{\rho_{k,1}^t}&{\rho_{k,2}^t}&{...}&{\rho_{k,n}^t}
\end{array}} \right]
\end{equation}
where ${z_{k,n}^t}$ is the available channel of the LEO satellites, and ${\rho_{k,n}^t}$ is the remaining visible time.
\item[$\bullet$]Action: $a_k^t$ represents the action of $k-th$ agent only served by one satellite, which is expressed as:
\begin{equation}
a_k^t = [\begin{array}{*{20}{c}}
0&0&{...}&1&{...}&0
\end{array}],{\rm{ }}\left\| a \right\| = 1
\end{equation}
\item[$\bullet$]Reward: $r_k^t(s_k^t,a_k^t)$ represents the instantaneous reward after agent $k$ at state $s_k^t$ executes action $a_k^t$ and interacts with the environment, which is expressed as $r_k^t(s_k^t,a_k^t)=$
\end{itemize}
\begin{equation}
\left\{ {\begin{array}{*{20}{c}}
{\begin{array}{*{20}{c}}
{ - 5\beta,{\rm{~if}}\;c_{k,n}^t =  = 0}
\end{array}}\\
{\begin{array}{*{20}{c}}
{ - \beta,{\rm{if}}(z_{k,n}^t \ge L) \vee (\vartheta_{k,n}^t  < {\vartheta _{threshold}})},
\end{array}}\\
{\rho _{k,n}^t{w_1} + \vartheta _{k,n}^t{w_2} + (L - z_{k,n}^t){w_3},{\rm{if}}~x_{k,n}^t = 1,x_{k,n}^{t - 1} = 1},\\
{\begin{array}{*{20}{c}}
{ - 0.5\beta + \rho _{k,n}^t{w_1} + \vartheta _{k,n}^t{w_2} + (L - z_{k,n}^t){w_3},\rm{else}}
\end{array}}
\end{array}} \right.
\end{equation}
In this reward function, we define four different cases, where $\beta$ is the upper bound of Case 3. In Case 1, it will get a penalty of -5$\beta$ when the user is not within the coverage of the satellites. In Case 2 where CINR does not meet the lowest threshold or network congestion happens, it will be punished by $-\beta$ to avoid this situation. In Case 3 where no handover
happens and the quality of the channel satisfies the reliable communication, it will get a positive value consisting of the
remaining visible time, CINR, and the available channel of the satellites. Otherwise, a penalty term of $-0.5\beta$ will be added to avoid large amounts of handover events in Case 4.

Soft Actor-Critic (SAC) is a variant of actor-critic algorithms, which includes a maximum entropy framework that encourages the agent to learn a more diverse policy by explicitly maximizing the entropy of the policy \cite{b16},
\begin{equation}
{\pi ^*} = \arg \mathop {\max }\limits_\pi  \sum\limits_{t = {t_0}}^{{t_U}} {{E_{({s_{t,}}{a_t})\sim{\tau _\pi }}}[{\gamma ^t}(r({s_t},{a_t}) + \alpha H(\pi (.|{s_t})))]} ,
\end{equation}
where $\alpha$ is the temperature parameters, $\gamma$ is the discount factor, ${\tau _\pi }$ is the distribution of trajectories induced by policy $\pi$, and $H(\pi (.|{s_t}))$ is the entropy of the policy $\pi$ at state $s_t$. It can be used for both continuous \cite{b16} and discrete \cite{b17} action spaces. For the discrete action spaces, since the action distribution can be recovered, we can delete the Monte Carlo estimate of the soft state-value function in the continuous environment and obtain the expectation directly \cite{b17}. Therefore, the new soft state-value calculation can be expressed as:
\begin{equation}
V({s_t}) = \pi {({s_t})^T}[Q({s_t}) - \alpha \log (\pi ({s_t}))].
\end{equation}
Also, the calculation of the temperature loss is adjusted to reduce the variance of the estimate which can be expressed as \cite{b17}:
\begin{equation}
J(\alpha ) = {\pi _t}{({s_t})^T}[ - \alpha (\log ({\pi _t}({s_t})) + \bar H],
\end{equation}
where $\bar H$ is the hyperparameter representing the target entropy. Moreover, the reparameterization trick \cite{b18} is deleted where the output of the policy network is combined with the vector sampled from the spherical Gaussian and the policy can be expressed as \cite{b17}:
\begin{equation}
{J_\pi }(\phi ) = {E_{{s_t}\sim D}}[{\pi _t}{({s_t})^T}[\alpha \log ({\pi _\phi }({s_t})) - {Q_\theta }({s_t})]]
\end{equation}
In the discrete state and action spaces, it also needs to train the soft Q-function to minimize the soft Bellman residual, which can be expressed as \cite{b17}:
\begin{equation}
\resizebox{1\hsize}{!}{${J_Q}(\theta ) = {{E_{({s_t},{a_t})\sim D}}[\frac{1}{2}{({Q_\theta }({s_t},{a_t}) - (r({s_t},{a_t}) + \gamma {E_{{s_{t + 1}}p({s_t},{a_t})}}[{V_{\overline \theta  }}({s_{t + 1}})]))^2}]}$}
\end{equation}
where $D$ is the replay buffer and $V_{\overline \theta}$ is estimated by the target network.

The concept of Nash Equilibrium (NE) is crucial in game theory as it provides a solution for analyzing the outcomes of strategic interactions involving multiple decision-makers. In \cite{b19}, \cite{b20}, NE has been combined with DRL to form Nash-DQN for mobile edge computing (MEC) to avoid overload. Nash-SAC has first been proposed for the formulated handover optimization problems, as is shown in Algorithm 1.
\begin{algorithm}
	\renewcommand{\algorithmicrequire}{\textbf{Input:}}
	\caption{Nash-SAC for LEO satellite handover optimization problem}
	\label{alg:1}
	\begin{algorithmic}[1]
		\REQUIRE Initialize network parameters $\overline {{\theta _1}}  \leftarrow {\theta _1},\overline {{\theta _2}}  \leftarrow {\theta _2},D \leftarrow \emptyset $\\
            \textbf{For} each episode do\\
            \quad obtain the current state information $s_{{k}}^{{t}}$\\
            \quad\textbf{For} section u=1:U do:\\
            \quad\quad select a Nash equilibrium action $a_{{k},Nash}^{{t}}$ by the game\\
            \begin{equation}
                a_{{k},Nash}^{{t}}\sim{\pi _\phi }(a_{{k}}^{{t}}|s_{{k}}^{{t}})
            \end{equation}
            \quad\quad Then, interact with the environment\\
            \quad\quad obtain the instantaneous reward:
            $r_{{k}}^{{t}}(s_{{k}}^{{t}},a_{{k}}^{{t}})$ \\
            \quad\quad and the next state information\\
            \begin{equation}
                s_{{k}}^{{t+1}}\sim p(s_{{k}}^{{t+1}}|s_{{k}}^{{t}},a_{{k}}^{{t}})
            \end{equation}
            \quad \quad Store them in $D$\\
            \begin{equation}
                D \leftarrow D \cup \{ (s_{{k}}^{{t}},a_{{k},Nash}^{{t}},r(s_{{k}}^{{t}},a_{{k}}^{{t}}),s_{{k}}^{{t+1}})\} 
            \end{equation}
            \quad\quad Sample random mini-batch of transitions from $D$ to training the Q-network;\\
            \quad \quad \textbf{For} each gradient step do\\
            \quad \quad \quad${\theta _m} \leftarrow {\theta _m} - {\lambda _Q}{\widehat \nabla _{{\theta _m}}}J({\theta _m}), m \in \{ 1,2\} $\\
            \quad \quad \quad$\phi  \leftarrow \phi  - {\lambda _\pi }{\widehat \nabla _\phi }{J_\pi }(\phi )$\\
            \quad \quad \quad$\alpha  \leftarrow \alpha  - \lambda {\widehat \nabla _\alpha }J(\alpha )$\\
            \quad \quad \quad$\overline {{Q_m}}  \leftarrow \tau {Q_m} + (1 - \tau )\overline {{Q_m}} , m \in \{ 1,2\} $\\
            \quad \quad \textbf{End} for\\
            \quad \textbf{End} for\\
            \textbf{End} for
	\end{algorithmic}  
\end{algorithm}

First of all, we initialize some hyperparameters such as the dimension of the hidden layers, the target entropy, and the learning rate. The replay memory $D$ is also emptied at first to store the past experience for the training, which owns a certain capacity. Next, we can obtain the current state information such as DL CINR of the current channel, the remaining visible time, and the available idle channel from the environment according to the time slot and type of users. Subsequently, we select a NE action based on the policy by the game. Afterwards, the agents interact with the environment which owns the global information of all LEO satellites and agents during the whole simulation period to obtain the instantaneous reward and the next-state information which will be stored in the replay buffer. Nash-SAC under the actor-critic scheme not only has the policy networks but also has the soft Q-value networks. Therefore, in the training process, when sampling a random mini-batch of transitions from the replay buffer, we need to update the policy network parameters by (15), the soft Q-value network parameters by (16), and the temperature parameter by (14) by performing a gradient descent step by the Adam optimizer. It is worth noting that Nash-SAC adopts double DQN which can avoid the over-estimate and Nash-SAC uses the soft update ($\tau  = 0.02$) every step to copy the parameters from the double soft Q-network to the double target network.

\section{Results AND Analysis}
\subsection{Parameters Setting}
In the LEO satellite network, we model the OneWeb satellite constellation from ephemeris data with global coverage and consider 10 aircraft, 10 UAVs, 10 eVTOLs, and 50 ground terminals in the scenario. Ground terminals and UAVs with a speed of 80 km/h are all uniformly distributed in the Stockholm area (about 45 km$^{2}$), the longitude span of which is 17.91 - 18.06 E, and the latitude span of which is 59.25 - 59.33 N. For eVTOLs with a speed of 240 km/h and aircraft with a speed of 900 km/h, the start points are also uniformly distributed in the Stockholm area (about 265 km$^{2}$), the longitude span of which is 17.91 - 18.20 E, and the latitude span of which is 59.25 - 59.65 N. For flying vehicles, 1/5 of them move towards Helsinki, 1/5 of them move towards Kiruna, 3/10 of them move towards Copenhagen, and 3/10 of them move towards Oslo. The simulation parameters in the DL communication are set in Table I, and the simulation period is 15 minutes.
\begin{table}[!ht]
\centering
\caption{Simulation Parameters of LEO Satellite Networks.}
~\\
\label{tab:sample-table-label}
\resizebox{1\columnwidth}{!}{
\begin{tabular}{|c|c|c|c|}
\hline
\textbf{} & \textbf{FVs} & \textbf{Ground terminal} & \textbf{LEO satellite}\\
\hline
Communication frequency& \multicolumn{3}{|c|}{18.5 GHz} \\
\hline
Total bandwidth& \multicolumn{3}{|c|}{250 MHz}  \\
\hline
Antenna type& \multicolumn{3}{|c|}{Parabolic} \\
\hline
Antenna size & 0.3/0.45 m& 0.54 m& 1 m\\
\hline
Noise temperature& \multicolumn{3}{|c|}{213.15-273.15 K} \\
\hline
Polarization isolation factor& \multicolumn{3}{|c|}{12 dB}\\
\hline
EIRP & \multicolumn{2}{|c|}{-} & 73.1 dBm\\
\hline
Minimum elevation angle& \multicolumn{3}{|c|}{15 degree}\\
\hline
${G_r}$/T& 14.2/15.0 dB & 15.4 dB&-\\
\hline
\end{tabular}
}
\end{table}

\subsection{Simulation Results and Analysis}
Maximum remaining service time (MRST)-based [13], maximum available channel (MAC)-based [12], maximum instantaneous signal strength (MIS)-based [11] and traditional Q-learning-based satellite handover strategies are selected as the benchmarks. Considering 8 available channels, as Fig. 2 shows, compared with the single criteria, traditional Q-learning, and Nash-DQN, Nash-SAC has better performance in average handover events, close to the lower limit. When using MIS, receiving signal strength is considered as the main criterion, which helps to prevent connection failures and maintain relatively high signal quality, but given the dynamic nature of the communication channel, rapid changes could result in unstable handover decisions and catastrophic network congestion. Moreover, the MAC-based strategy considers more idle available channels to avoid overloading the satellites, while it neglects considering relative positions with the LEO satellite and the quality of the channels which leads to frequent handovers. In contrast, MRST provides a lower limit of average handovers but can not deal with network congestion as is shown in Fig. 3 For the traditional Q-learning method, it sinks into the local minimum with insufficient exploration even if with the $\epsilon$-greedy policy, thereby it has still a high blocking rate and high average handovers. Compared with traditional Q-learning, Nash-SAC greatly optimizes the efficiency of traditional algorithms and can not only reduce the number of handovers by 16\% but also improve the blocking rate by 18\%, which proves the efficiency of the strategy when combined MARL with the game theory. Compared with Nash-DQN, Nash-SAC under the actor-critic scheme includes the maximum entropy framework with the stochastic policy, which encourages more exploration in the training process, and the soft temperature parameter can also balance the exploration and exploitation well. Moreover, NE encourages cooperation over the fierce competition where Nash-DQN and Nash-SAC may choose other communication channels which may have lower CINR, avoiding degrading the whole system, while Nash-SAC is superior to Nash-DQN.
\begin{figure}[t]
\centerline{\includegraphics[width=2.8in]{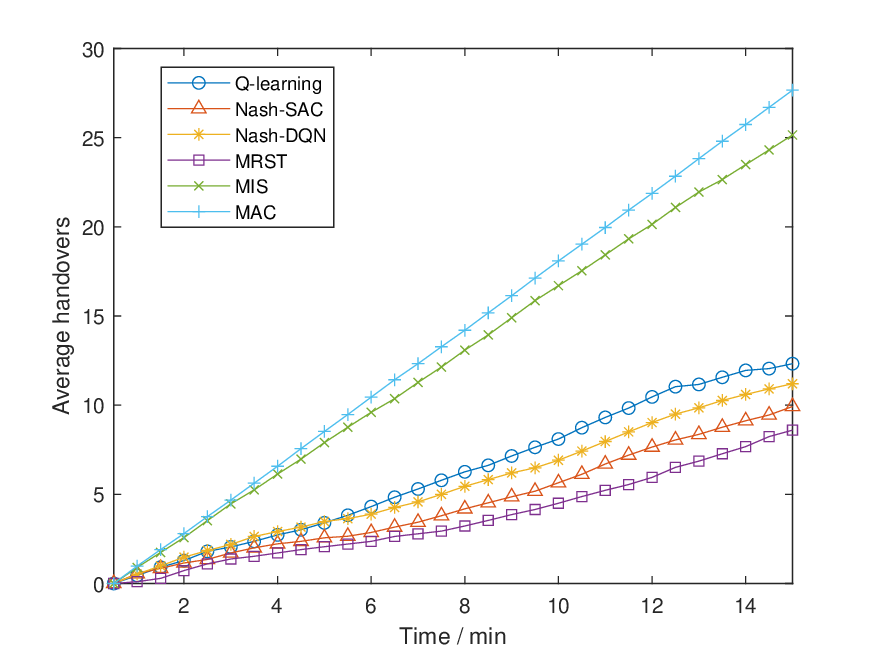}}
\caption{Number of accumulative average handovers versus time.}
\label{fig}
\end{figure}
\begin{figure}[t]
\centerline{\includegraphics[width=2.8in]{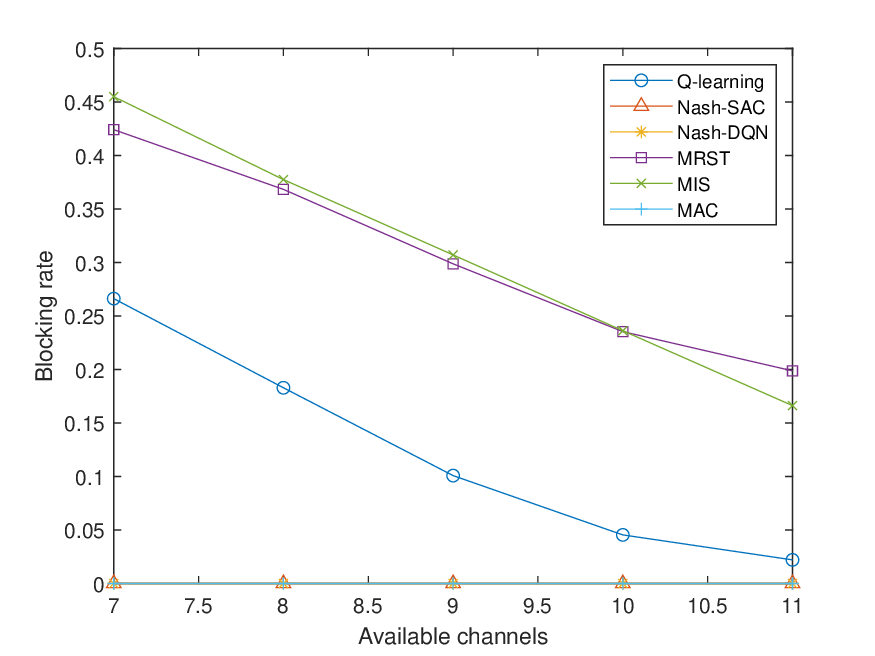}}
\caption{Blocking rate versus available channels when using different methods.}
\label{fig}
\end{figure}
\begin{figure}[b]
\centerline{\includegraphics[width=2.8in]{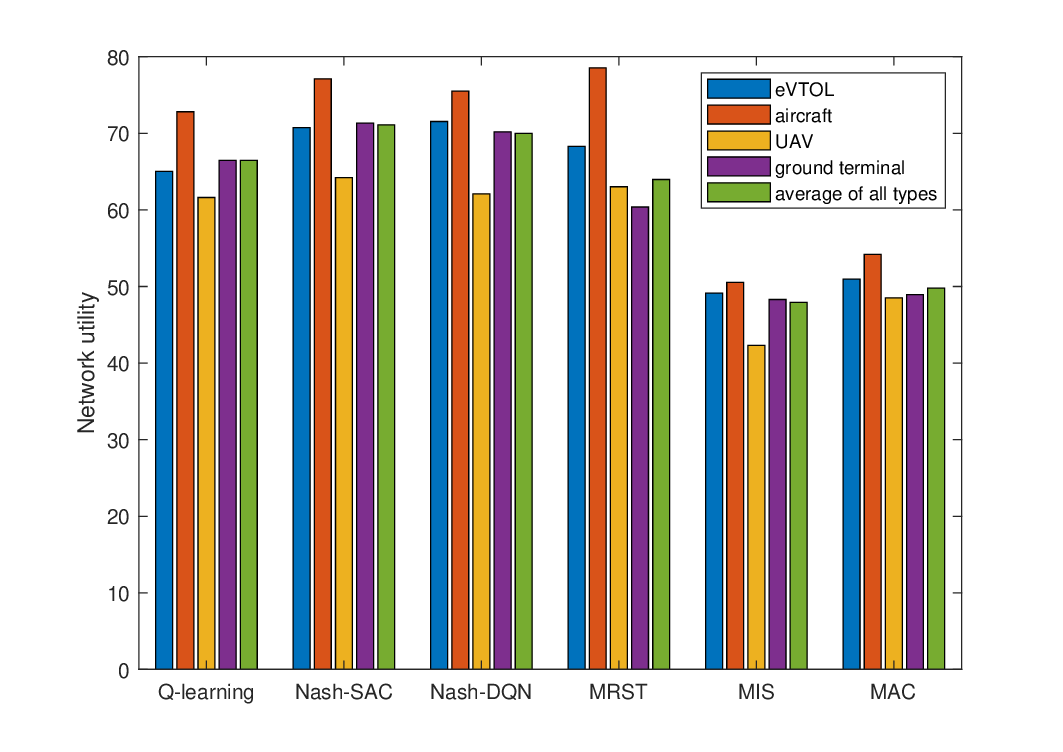}}
\caption{Network utility for different types of users when using different methods.}
\label{fig}
\end{figure}
\begin{figure}[t]
\centerline{\includegraphics[width=2.8in]{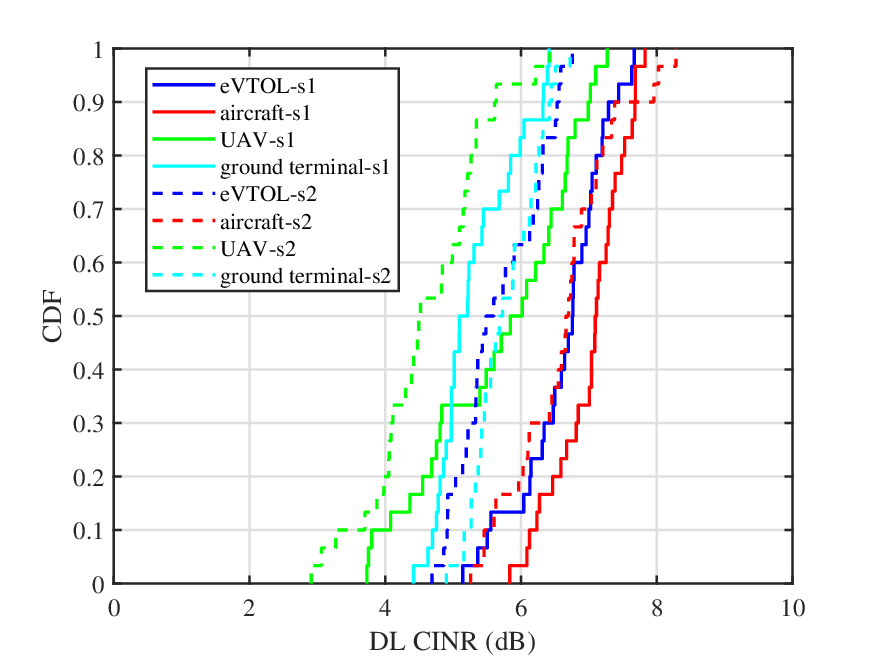}}
\caption{CDF of DL CINR for different types of users when using Nash-SAC.}
\label{fig}
\end{figure}

Considering the second optimization problem, from Fig. 4, we can conclude that Nash-SAC has the best performance with $71.1$ of network utilization, outweighing Nash-DQN ($70.0$), Q-learning ($66.5$), MRST ($64.0$), MAC ($49.8$) and MIS ($47.9$). When comparing different FVs and the ground terminals, the aircraft obtain the best communication channel owing to their high altitude with low signal attenuation and the line-of-sight (LOS) path. The ability of UAV and eVTOL to obtain high-quality channels is inferior to the aircraft, and their network utilization efficiency is close to the ground terminals. Although the ground terminal has lower power density in terms of such a long distance, it is configured with a larger antenna size with higher gain. Due to the size of UAVs, it is not suitable to be equipped with such a large antenna. In addition, the number of ground terminals is much higher than different FVs which may cause resource competition conflicts. Therefore, it is essential to give priority to FVs in some cases for safety problems.

Fig. 5 shows the cumulative distribution function (CDF) of DL CINR when using Nash-SAC in scenario s1 (priority to FVs) and scenario s2 (no priority).  It can be concluded that for s1, the channel quality of FVs has been significantly improved. Among different types of users, the improvement in DL CINR for eVTOLs and UAVs is particularly significant, with the cost in turn, and the channel quality of the ground terminals has slightly decreased but it is tolerable for them to maintain a CINR level between 5-6 dB.

\section{Conclusions}
In this paper, the realistic LEO satellite system model is constructed from the ephemeris data. Different FVs, ground terminals, and LEO satellites constitute networks, and two mathematical optimization problems are formulated considering the comprehensive criteria, including the remaining visible time, signal quality CINR, and the available idle channel. To solve these problems, a novel satellite handover strategy based on Nash-SAC which combines the MARL and game theory has been proposed. Moreover, simulation results with different benchmarks are analyzed, showing that the proposed strategy achieves the best performance to provide reliable and robust connectivity with the lowest satellite handovers, lowest blocking rate, and the highest network utility. When the number of ground terminals is enough high, the priority for the FVs also works to avoid the disadvantages state in fierce conflicts, which proves the feasibility and robustness of the proposed handover strategy.

\section*{Acknowledgment}
This work was supported in part by the CELTIC-NEXT Project, 6G for Connected Sky (6G-SKY), with funding received from Vinnova, Swedish Innovation Agency.

\vspace{12pt}

\end{document}